\documentclass[showpacs,floatfix,eqsecnum,aps,nofootinbib]{revtex4}  
\usepackage{epsfig}

\begin{document}

\title{Littlest Higgs Model and Unitarity Constraints}

\author{Namit Mahajan}
\email{nmahajan@mri.ernet.in}
\affiliation{	{\em Harish-Chandra Research Institute,} \\
	 {\em Chhatnag Road, Jhunsi, Allahabad - 211019, India.}}

\def\be{\begin{equation}}
\def\ee{\end{equation}}
\def\bea{\begin{eqnarray}}
\def\eea{\end{eqnarray}}
\def\ba{\begin{array}}
\def\ea{\end{array}}

\begin{abstract}
The Littlest Higgs Model is constrained using the tree-level
perturbative arguments, similar to those employed to obtain the Higgs mass
bound. It is found that the perturbative unitaity violations set in at
energy scales $\sim 1.5\chi f$, where
$f$ is the analog of pion decay constant and sets the scale for the
masses in the theory and also the strong interaction scale and $\chi$
is a dimensionless parameter which together with $f$ sets the mass scale of
heavy scalars in the theory.
\end{abstract}
\maketitle

\begin{section}{Introduction}
\indent The Standard Model (SM) of particle physics seems to agree with almost
all the experimental data and has been tremendously successful in
predicting particle properties and interactions. However, till date
the mechanism of electroweak symmetry breaking is a mystery. The most
popular one,
the Higgs mechanism, requires fundamental scalar(s) in the theory,
though none has been observed till date (for a review of the basic
ideas involved in Higgs mechanism, some of the un-answered questions
and remedies, see \cite{sally}). Indirect evidences point to
a light Higgs boson \cite{precision}. 
A fundamental scalar in the theory implies that
the mass term for this scalar particle will receive large quadratic
corrections proportional to the largest scale, which generically would
be identified with the Planck scale or some other very high
fundamental scale in the full theory, including gravity finally. 
This leads to a large disparity, called the {\it hierarchy problem},
between the electroweak scale, fixed by the Fermi constant,
 and this high scale. To circumvent this problem, many solutions
have been propsed and all of them have invited intense activity.
Supersymmetry (for an introduction of basic ideas see \cite{susy}) 
is the most attractive of these proposals. The
symmetry between the fermions and bosons renders the quadratic
corrections to the Higgs mass contribution small provided the mass
splitting between the SM particles and their super-partners is not too
large, further implying that the supersymmetry breaking scale is not
to large compared to the electroweak scale. 
However, there is no compelling reason to believe the existence
of low energy supersymmetry. The other possiblity which has attracted a
lot of attention in the last couple of years is the idea that the four
dimensional Planck scale is just a derived quantity while the
fundamental scale of gravity can be as low as TeV \cite{ed}. 
This idea required the existence of one or more, compact or non-compact,
 spatial extra dimensions.\\ 
\indent Almost all the alternative scenarios invoked in 
order to circumvent the hierarchy
problem have their own set of drawbacks or short comings. In
particular, the fact that they require the introduction/inclusion of a
large number of new, and sometimes very exotic, particles makes them
very speculative and in some sense reduce the predictive power of the
theory as a whole. \\
\indent Given the state of affairs, it seems very reasonable to look
for other possible ways of getting across this problem. Recently,
there has been a revived interest in the idea of the Higgs boson being
a Pseudo-Goldstone Boson (PGB) \cite{pgb}.
Exploiting this idea, Arkani-Hamed etal \cite{nima1} 
have proposed a minimal model, the Littlest Higgs model,
where there are no elementary scalars in the theory, thus avoiding the
issue of quadratic divergence right from the start. Motivated by the
dimensional deconstruction ideas \cite{deconstruct}, it was shown that
the little Higgs models can be successfully realised \cite{lh}
with the basic
feature that the quadratic divergences to the Higgs mass are absent to
one loop level. In the little Higgs models, very similar to models
with dynamical symmetry breaking (for details see for example 
\cite{dynamical}), there are no
elementary scalars. An attractive and perhaps plausible way to obtain
these scalars is to imagine a scenario where 
some strong dynamics is responsible for binding the
elementary fermions into composite scalars, which subsequently 
play the role of the Higgs boson. 
At the scales relevant for such strong dynamics, the
fermions condensate and break the ``strong'' group, thereby
giving rise to Goldstone bosons, the details and number of which are
model dependent. The situation is very
similar to chiral symmetry breaking in QCD \cite{chiral}. However, it
should be kept in mind that this might just be one of the possible ways
of achieving the goal and that other possibilities are not 
ruled out apriori.\\ 
\indent In the model considered in \cite{nima1}, it is assumed that
the theory has an $SU(5)$ global symmetry which is spontaneously
broken to $SO(5)$ yielding a large number of Goldstone bosons. This is
assumed to happen at some large scale where the fermions form
condensates and break the symmetry. The model assumes that there are
two copies of $SU(2)\times U(1)$ (direct product groups) 
embedded in $SU(5)$ which break to $[SU(2)\times U(1)]_{SM}$ at the
same scale as that of the global symmetry breaking. The presence of a
larger symmetry group ensures that the Higgs mass is approximately
protected and therefore avoids the problem of large quadratic
divergences upto one loop level. The electroweak symmetry breaking is
triggered via the Coleman-Weinberg mechanism. \\
\indent Several variations of this idea have
also been studied \cite{variations}.
Some of the phenomenological aspects of such scenarios have
also been explored \cite{pheno}. 
In this note we constrain the minimal model \cite{nima1}
 by invoking arguments based on tree level perturbative
unitarity of the the scattering amplitudes.
\end{section} 
%%%%%%%%%%%%%%%%%%%%%%%%%%%%%%%%%%%%%%%%%%%%%%%%%%%%%%%%%%%%5
\begin{section}{Littlest Higgs Model: A quick look}
\indent Consider the littlest Higgs model \cite{nima1}
based on non-linear sigma
model describing $SU(5)/SO(5)$ symmetry breaking, originating due to
vacuum expectation value (VEV) $\sim {\mathcal{O}}(f)$
 of a symmetric tensor. By dimensional
analysis \cite{dim}, 
this implies that the symmetry breaking scale $\Lambda \sim
4\pi f$. Let us further choose a convenient basis in which the VEV
points in the following direction, labelled by $\Sigma_0$
\be
\Sigma_0 = \Bigg(\ba {clll}
                  & &1_{2\times 2} \\
                  &1& \\ 
                  1_{2\times 2}& & \ea \Bigg)
\ee
Further, if $X^a$ are the broken generators, then the Goldstone bosons
($\pi^a$) can be parameterized by the non-linear sigma field as
\be
\Sigma(x) = e^{2i\Pi^T/f}\Sigma_0
\ee
where $\Pi = \pi^aX^a$. The breaking $SU(5) \to SO(5)$ yields fourteen
Golstone bosons out of which four are absorbed by the gauge fields
(the heavy gauge fields in the model)
to acquire masses when the $[SU(2)\times U(1)]^2$ sub-group is gauged
to $SU(2)_L\times U(1)_Y$. In terms of the remaining ten massless
scalars, the Goldstone boson matrix $\Pi$ can be written as
\be 
\Pi = \Bigg(\ba{clll}
            0& \frac{h^{\dag}}{\sqrt{2}}&\phi^{\dag}\\
            \frac{h}{\sqrt{2}}&0&\frac{h^*}{\sqrt{2}}\\
            \phi&\frac{h^T}{\sqrt{2}}&0 \ea \Bigg)
\ee 
where $h$ is a complex doublet under the SM gauge group while $\phi$
is a complex triplet forming a $2\times 2$ symmetric tensor. The
breaking of $[SU(2)\times U(1)]^2$ to the SM gauge group means that
the four vector bosons acquire masses of the order of $f$ and
therefore are heavy. The non-linear sigma model Lagrangian (to the leading
order) reads
\be
{\mathcal{L}}_{\Sigma} \sim \frac{f^2}{4}Tr\vert{\mathcal{D}}_{\mu}
\Sigma\vert^2
\ee
where ${\mathcal{D}}_{\mu}$ is the relevant covariant derivative
including the gauge fields and couplings corresponding to both the
copies of $SU(2)\times U(1)$:
\be
{\mathcal{D}}_{\mu}\Sigma = \partial_{\mu}\Sigma -
i\sum_{j=1}^2[g_jW_j^a(Q_j^a\Sigma + \Sigma{Q_j^a}^T) + 
g_j^{\prime}B_j(Y_j\Sigma + \Sigma Y_j)]
\ee
where $g_j$, $g_j^{\prime}$ are the gauge couplings, $W_j$, $B_j$ are
the gauge bosons and $Q_j^q$ and $Y_j$ are the generators of gauge 
transformations. The gauge couplings break the global symmetry
explicitly. However, no single interaction breaks the complete
symmetry and therefore the Higgs mass is partially protected.
Expanding $\Sigma$ about $\Sigma_0$
gives the linearized theory and breaks the local gauge symmetry to the SM
gauge symmetry. The massless gauge bosons are identified with the SM
gauge bosons. In what follows, we closely follow the notations and
conventions chosen in Han etal. \cite{pheno}. 
The breaking of the local gauge symmetry to the SM gauge group
generates masses for four gauge bosons. Linear combinations of $W_j$
and $B_j$ are respectively constructed such that one set becomes
massive while the other set remains massless and is identified with
the SM gauge bosons. Also, in terms of the gauge couplings $g_j$ and
$g_j^{\prime}$ the mixing angles are defined as:
\be
s = \frac{g_2}{(g_1^2+g_2^2)^{\frac{1}{2}}} \hskip 1.5cm
s^{\prime} = \frac{g_2^{\prime}}{({g_1^{\prime}}^2+
{g_2^{\prime}}^2)^{\frac{1}{2}}}
\ee
The SM gauge couplings are finally identified as 
\be 
g = g_1s = g_2c \hskip 1.5cm g^{\prime} = g_1^{\prime}s^{\prime} = 
g_2^{\prime}c^{\prime}
\ee
Let $v$ and $v^{\prime}$ denote the Higgs and triplet VEVs
respectively. Expanding the Lagrangian in terms of the fields $h$ and
$\phi$, and minimising the potential, we arrive at an expression for
the triplet mass ($m_{\Phi}$). 
By demanding that that the triplet mass squared is a
positive definite quantity, the two VEVs get related in the following
fashion:
\be
\frac{{v^{\prime}}^2}{v^2} \le \frac{v^2}{16f^2} \label{vevrel}
\ee
Therefore, we can now trade off $v^{\prime}$ for $v$ and $f$.\\ 
\indent In the fermionic sector, one of the main concerns is to
generate a large top-Yukawa coupling and at the same time ensuring that
there are no large quadratic corrections due to top quark. In the
present scenario, this is accomplished by introducing a pair of
coloured Weyl fermions labelled - $\tilde{t}$ and $\tilde{t^c}$ in
addition to the third family doublet comprising of bottom and top
quark. Expanding the relevant piece to first order in $h$ and
rearranging by suitably combining the different fields yields large
mass $\sim {\mathcal{O}}(f)$ 
for the vector like fermion and generates the desired 
top-Yukawa coupling. In this minimal framework,
no additional fields are introduced for any of the other fermions. \\ 
\indent Finally, the electroweak symmetry breaking is induced via the
Coleman-Weinberg mechanism \cite{coleman} and therefore the 
quadratic corrections to Higgs mass are absent to one loop level.  
Therefore, below the scale $f$, this minimal
model has the particle content that is identical to SM and nothing
else. Moreover, as the scale $f$ is approached, there is a minimal set
of heavy fields that gets introduced. This is in some sense a nice
economical feature of the model. Therefore, in this scenario, at least
two operators are needed to completely break the global symmetry that
protects the Higgs mass, forbidding quadratic divergences at one loop
level. The Higgs mass gets quadratically corrected at the two loop
level and hence is much smaller than $\Lambda$. We thus have
a situation : $\Lambda > f > v$ such that the successive hierarchies
are small. Above the global symmetry breaking
scale, we require
a UV completion mechanism for the theory. However, for the present
study, we don't need to be bothered about the details of such a
mechanism.
\end{section}
%%%%%%%%%%%%%%%%%%%%%%%%%%%%%%%%%%%%%%%%%%%%%%%%%%%%%%%%%%%%%%
\begin{section}{Unitarity Constraints}
\indent The question generally asked while
relying on perturbative calculations is regarding the energy scales
upto which the theory can be treated perturbatively. Put in another
form, it is desirable to have an estimate of the
cut-off scale governing the low energy dynamics in the perturbative
fashion. It is well known that such an issue can be addressed to a
fair degree of accuracy on the arguments based on tree level
perturbative requirements of the scattering amplitudes. In SM, it is
known that the Higgs boson plays a crucial role in restoring the
unitarity of longitudinal gauge boson scattering amplitudes and
this in turn gives a bound on the Higgs mass \cite{lee}.\\
\indent We use these very ideas to get an idea of the
scale upto which we can naively treat little Higgs models
perturbatively, without bothering about the strong dynamics setting
in. To this effect, we study scattering of longitudinal vector bosons
in this model. We expect that the longitudinal gauge boson
interactions should bring out any features of the strong interactions
because the {\it equivalence theorem} \cite{equi} in its simplest and
navive form states that that for
energies much larger than the vector boson masses, the scattering
of the longitudinal gauge bosons can be viewed as scattering of the
Goldstone bosons. This simplifies the calculations
drastically. In practice, the polarization vectors for the
gauge bosons are taken to be
\[
\epsilon^{\mu}(p_V) = \frac{p_V^{\mu}}{M_V} + {\mathcal{O}}(M_V^2/E^2)
\]
where $p_V$ and $M_V$ are the vector boson momentum and mass
respectively and $E>>M_V$ characterises the typical energy scale involved
in the scattering process. It may be argued that this naive form of 
the equivalence theorem is not
applicable when dealing with effective theories or chiral Lagrangians
because of the fact that effective theories imply the low energy limit
of the full theory while equivalence theorem implies that we have to
take the high energy limit. These two limits have to be simultaneously
taken and it may not always be possible to successfully implement this
procedure. Also, in the case of effective theories, since the energy
scales are restricted to a certain value, there is an infinite series
of higher dimensional operators in the low energy theory and in such a
situation the validity of the naive form of equivalence theorem is in
question. These aspects have been discussed in \cite{effective}. Also,
we are interested in the longitudinal gauge boson scattering
amplitudes. But under a general Lorentz transformation, the
longitudinal and transverse components of the gauge field can mix and
therefore the replacement of the longitudinal components by their
associated Goldstone bosons can lead to Lorentz non invariant results
\cite{he}. However, the requirement, $E >> M_V$ circumvents such
difficulties and defines the suitable Lorentz frame(s).
Also, this ensures that the
${\mathcal{O}}(M_V^2/E^2)$ terms are actually small and can be safely
neglected. Furthermore, while considering the effective theory
amplitudes, the correct energy regime to be employed to
use the equivalence theorem in its simplest form and without bothering
about other corrections (except for the renormalization factors
arising at loop level), is $M_V<<E<4\pi\epsilon$ where $\epsilon$ is
the energy scale that fixes the gauge boson masses \cite{applicability}.\\  
\indent In the present case, the model predicts a very light Higgs,
whose mass is protected by some approximate symmetries upto one loop
level and a complex triplet of scalars. Expanding the sigma model
Lagrangian to relevant order, it is clear that there are additional
corrections to the usual SM vertices. Also, the scattering processes
will receive additional contributions from the new vertices and
particles present. In the SM, the Higgs boson restores the unitarity
of the amplitudes and there is a very delicate cancellation between
various diagrams that effects such a beautiful restoration. However,
in the case when there are deviations, however small, from the exact
SM vertices and more so when there are additional particles in the
theory, such a cancellation may not be operative so
effectively. Following
the expanded Lagrangian as in Han etal \cite{pheno}, 
it is easy to convince that
the pure SM pieces will be nicely behaved. All the bad behaviour
contributions either stem out due to the presence of the triplet of
scalars or some residual contribution from the corrections to SM
vertices. In the notation of Han etal \cite{pheno}, 
we make the following allowed phenomenological choices,
\be
c = s \hskip 1cm c^{\prime} = s^{\prime} \hskip 1cm
\frac{v^{\prime}}{v} = \frac{v}{\chi f} \label{choices}
\ee
with $\chi>4$ such that Eq.(\ref{vevrel}) is satisfied. We choose
different values of the parameter $\chi$ and investigate the effects
on unitarity violation.\\
\indent We consider the coupled longitudinal vector boson scattering
amplitudes (keeping in mind the
discussion above regarding the validity and use of equivalence theorem
in chiral/effective theories)
which on using these choices result in the
following $J=0$ partial wave amplitudes:
\bea 
{\mathcal{M}}(W_LW_L\to W_LW_L) &=& \Bigg(-\frac{ig^2}{64\pi M_W^2}\Bigg)
\Bigg[f_{WWH}^2\Bigg(2m_H^2 + \frac{m_H^4}{s-m_H^2} -
  \frac{m_H^4}{s}\ln(1+s/m_H^2)\Bigg) \nonumber \\
&+& f_{WW\Phi^0}^2\Bigg(2m_{\Phi}^2 + \frac{m_{\Phi}^4}{s-m_{\Phi}^2 +
  im_{\Phi}\Gamma_{\Phi}} -
  \frac{m_{\Phi}^4}{s}\ln(1+s/m_{\Phi}^2)\Bigg) \\ \nonumber
  &+& f_{WW\Phi^{--}}^2\Bigg(m_{\Phi}^2 - \frac{s}{2} -
  \frac{m_{\Phi}^4}{s}\ln(1+s/m_{\Phi}^2)\Bigg)\Bigg]
\eea
\bea 
{\mathcal{M}}(Z_LZ_L\to Z_LZ_L) &=& \Bigg(-\frac{ig^2}{64\pi M_W^2}\Bigg)
\Bigg[f_{ZZH}^2\Bigg(3m_H^2 + \frac{m_H^4}{s-m_H^2} -
  2\frac{m_H^4}{s}\ln(1+s/m_H^2)\Bigg) \nonumber \\
&+& f_{ZZ\Phi^0}^2\Bigg(3m_{\Phi}^2 + \frac{m_{\Phi}^4}{s-m_{\Phi}^2 +
  im_{\Phi}\Gamma_{\Phi}} -
  2\frac{m_{\Phi}^4}{s}\ln(1+s/m_{\Phi}^2)\Bigg) \Bigg]
\eea
\bea
{\mathcal{M}}(W_LW_L\to Z_LZ_L) &=& \Bigg(-\frac{ig^2}{64\pi M_W^2}\Bigg)
\Bigg[\Bigg(m_H^2 + \frac{m_H^4}{s-m_H^2}\Bigg) + 
(f_{WWH}-1)(f_{ZZH}-1)\frac{s^2}{s-m_H^2} \nonumber \\
&+& 
f_{WW\Phi^0}f_{ZZ\Phi^0}\frac{s^2}{s-m_{\Phi}^2 +
  im_{\Phi}\Gamma_{\Phi}}
\\ \nonumber 
&+& f_{WZ\Phi^-}^2\Bigg(2m_{\Phi}^2 - s -  
2\frac{m_{\Phi}^4}{s}\ln(1+s/m_{\Phi}^2)\Bigg) \Bigg]
\eea
where 
\[
f_{WWH} = 1 - \frac{v^2}{3f^2} - \frac{s_0^2}{2} - 2\sqrt{2}
s_0\frac{v^{\prime}}{v} \hskip 1cm
f_{ZZH} = 1 - \frac{v^2}{3f^2} - \frac{s_0^2}{2} + 
4\sqrt{2}s_0\frac{v^{\prime}}{v}
\]

\be
f_{WW\Phi^0} = s_0 - 2\sqrt{2}\frac{v^{\prime}}{v}
\hskip 1.5cm f_{WW\Phi^{--}} = 4\frac{v^{\prime}}{v}
\ee

\[
f_{ZZ\Phi^0} = s_0 - 4\sqrt{2}\frac{v^{\prime}}{v}
\hskip 1.5cm  f_{WZ\Phi^-} = 2\frac{v^{\prime}}{v}
\]
with 
\be 
s_0 \simeq 2\sqrt{2}\frac{v^{\prime}}{v}
\hskip 1.5cm
m_{\Phi} = \frac{\sqrt{2}m_H f}{v}\frac{1}{[1-(4v^{\prime}f/v^2)^2]^
{\frac{1}{2}}} \label{mphimass}
\ee
where the decay width of the triplet, $\Gamma_{\Phi}$ has been
explicitly included in the above amplitudes.\\
\indent In writing ${\mathcal{M}}(W_LW_L\to Z_LZ_L)$, we've
deliberately separated the pure SM (the first two terms in the
parenthesis) and the new contributions. It is clear that
$f_{WWH},~f_{ZZH} \to 1$ while 
$f_{WW\Phi^0}, ~f_{ZZ\Phi^0},~f_{WW\Phi^{--}},~f_{WZ\Phi^-} \to 0$,
when $v/f \to 0$ and the
$J=0$ amplitudes correspondingly reduce to the pure SM amplitudes.
Also to be noted is the fact that with these choices of the parameters
Eq.(\ref{choices}), there are no contributions to $WW$ scattering
amplitude due to heavy gauge bosons. \\  
\indent The coupled channels form a $2\times 2$ matrix. This matrix is
diagonalised and the unitarity constraint is imposed on the larger of
the two eigenvalues ie if $\lambda$ denotes the larger eigenvalue then
the unitarity constraint reads
\be
\vert Re(\lambda)\vert \le \frac{1}{2} \label{unieq}
\ee
giving an inequality in terms of $m_H$, $\sqrt{s}$,
$f$, $\Gamma_{\Phi}$ and the parameter $\chi$.
For this study, we choose $m_H=115$ GeV as
indicated by the LEP results \cite{lep}. This considerably simplfies
our discussion as well because the pure SM amplitudes are well behaved
at high energies and respect the unitarity conditions for a light
Higgs. \\
\indent Furthermore, in the notation of
\cite{pheno} and as indicated in Eq.(\ref{mphimass}), $s_0$ is very
close to $2\sqrt{2}v^{\prime}/v$ and therefore in such a situation
$f_{WW\Phi^0} \to 0$. To investigate the implications of the unitarity
condition, we solve Eq.(\ref{unieq}) for different choices of the
parameters. The results are shown in Table 1.\\
\begin{table}[ht]
\begin{center}
\begin{tabular}{|c|c|c|c|}\hline
$\chi \rightarrow$ & 5&10&25\\
f (TeV) $\downarrow$&&&\\ \hline
0.5&4&8.5&14.5\\ \hline
1&8.2&16.5&44\\ \hline
2&16.5&32.2&83.2\\ \hline
\end{tabular}  
\end{center}
\caption{Unitarity violation scale (in TeV) for various values of $f$
and $\chi$}
\end{table}
%%%%%%%%%%%%%%%%%%%%%%%%%%%%%%%%%%%%%%%%%%%%%%%
For a light Higgs as we are considering, the SM contributions
asymptotically approach a constant value and therefore do not bother
us with any significant effects. Also to be noted is that in the regime $s >>
m_{\Phi}^2$ ie when all terms of the form $m_{\Phi}^2/s$ and higher powers
of the same are safely neglected, 
the amplitudes scale as $s/(f^2 \chi^2)$ which is clearly depicted in
Table 1 as well. In fact, it is not hard to conclude that the scale of
unitarity violation sets in at
\be
E_{unitarity} = {\sqrt{s}}_{critical} \sim (1.5-1.7)~\chi f 
\ee 
\indent Further, to
ensure that the results are not faked by any resonant behaviour, 
we have chosen $\Gamma_{\Phi}=50GeV$. This may seem a
bit artificial, but for the present study, it serves the purpose of
keeping the amplitudes well below the unitarity limits even when the
center of mas energy is close to the triplet mass. Making
$\Gamma_{\Phi}$ smaller will only result in a very sharp and narrow peak.\\
\indent For comparison, we list the triplet
mass for various $\chi$ values:
\[
m_{\Phi}(\chi=4) = \infty \hskip 1cm m_{\Phi}(\chi=5) \sim 1.1f
\]
\[
m_{\Phi}(\chi=10) \sim 0.71f \hskip 1cm m_{\Phi}(\chi=25) \sim 0.66f
\]
Clearly, increasing $\chi$ has the effect of decreasing the triplet
mass and as can be seen from Table 1, this has the effect of making
the scale of unitarity violation larger, as expected. Letting
$\chi$ take a large value results in a much larger scale at which
the unitarity violations show up. However, this variation of the
triplet mass with $\chi$ is not found for very large values of $\chi$ for
which the triplet mass becomes practically independent of the $\chi$
value, as can be seen from Eq.(\ref{mphimass}) and using the relation
between $v$ and $v^{\prime}$.
From the present analysis, it is
quite transparent that ${\sqrt{s}}_{critical} > 2f$ for all the
allowed values. This has the simple interpretation that, modulo
multiplicative factors of order unity, the theory under consideration
respects perturbative unitarity upto these scales. In particular, we
can safely conclude that the effective theory description is
valid upto scales ${\mathcal{O}}(2m_{\Phi})$ and in fact, much beyond. 
This is in agreement
with the general results and arguments advocated in \cite{hasenfratz}
for the validity and reliability of effective theory descriptions.\\ 
\indent These limits are not changed significantly if we
vary the Higgs mass even upto 150 GeV. For $\chi=5$,
Eq.(\ref{mphimass}) implies that the mass of the heavy scalars
(triplet) $\sim 1.1~f$ and mass of the heavy top $\sim 1.5~f$.
Also, the masses of the heavy $SU(2)$ gauge bosons are of the order of
the triplet mass. The heavy $U(1)$ gauge boson 
is not really very heavy, $\ge 0.15~f$, and can cause 
observable effects in the precision
measurements. Park and Song \cite{pheno} have pointed out
that even a 200 GeV heavy $U(1)$ gauge boson gives a negligible
contribution to muon magnetic moment.\\
\indent In this study we have been concerned with the unitarity of the
lighter of the two sets of vector bosons predicted by the littlest
Higgs model. The unitarity limits can change if the
heavier set scattering amplitudes are also accounted for. Nonetheless,
the limits obtained from the lighter gauge boson scattering amplitudes
are going to be much more stringent and we therefore stick to them in
the present note.  
\end{section}  
%%%%%%%%%%%%%%%%%%%%%%%%%%%%%%%%%%%%%%%%%%%%%%%%%%%%%%%%%%%%%%
\begin{section}{Conclusions}
\indent In this note, we have studied the tree-level perturbative
unitarity constraints on the littlest Higgs model. It is found that
demanding the perturbative validity of the SM, in the minimally
modified form, we obtain the unitarity violating scale $E_{unitarity}
\sim 1.5~\chi f$ with $\chi > 4$ always. For the limiting case, when
$\chi$ is marginally greater than $4$, $E_{unitarity}\sim (6-7) f$,
smaller than $4\pi f$, the scale of strong dynamics. For larger values
of $\chi$, the requirement that $M_V << E << 4\pi f$ is not met and
therefore, the equivalence theorem cannot be used in such a simple
manner. To avoid such difficulties and to get physically meaningful
results, it is therefore judicious to have $\chi \sim 5$ so as to
satisfy the above requirement.
In this study, we have only considered the longitudinal gauge boson
scattering. Perhaps, the limits obtained using the present analysis
can be pushed a bit if we consider other channels involving the Higgs
boson and top quark as well. Going beyond the tree-level would
introduce renormalization corrections to the equivalence theorem and
can further modify the present limits.\\
\indent The limits on $f$ obtained in the present
note agree with the general expectation based on other
phenomenological explorations \cite{pheno}. However, one important
point that can cause some modification in these values is the fact
that there is some amount of hidden uncertainity pertaining to the
specific UV completion mechanism for such a model. Nevertheless, we
believe that such an effect will only introduce a multiplicative
correction factor ${\mathcal{O}}(1)$. Also, we have chosen very
  specific values for the various mixing angles in our study. Changing
  these values does not significantly affect our results as the
  corrections introdued due to such changes will be small.
Only detailed studies can make very specific and
  accurate predictions about these mixing parameters which can, in
  principle, make significant effects in some other sectors.\\
\indent In conclusion, we like to mention that with the chosen
parameter set, the unitarity bound on the
parameter $f$ which sets the mass scale for the heavy particles and also
the scale of the strong dynamics, is in gross agreement with the
phenomenological constraints. Similar resulst can be expected for any
other variant of this minimal model considered here.
\end{section}  
%%%%%%%%%%%%%%%%%%%%%%%%%%%%
\acknowledgements 
I would like to thank Spencer Chang for pointing out a numerical error
in one of the expressions.
%\pagebreak

%%%%%%%%%%%%%%%%%%%%%%%%%%
%
%%%%%%%%%%%%%%%%%%
\end{document}